\documentclass[sigconf]{acmart}
\AtBeginDocument{%
  }

\setcopyright{acmlicensed}
\copyrightyear{2018}
\acmYear{2018}
\acmDOI{XXXXXXX.XXXXXXX}
\acmConference[Conference acronym 'XX]{Make sure to enter the correct
  conference title from your rights confirmation email}{June 03--05,
  2018}{Woodstock, NY}
\acmISBN{978-1-4503-XXXX-X/2018/06}

\usepackage{amsmath,amssymb,amsfonts}
\usepackage{enumitem}
\usepackage{bbm}
\usepackage{algorithmic}
\usepackage{graphicx}
\usepackage{textcomp}
\usepackage{xcolor}
\usepackage{booktabs}
\usepackage{adjustbox}
\usepackage{multirow}
\usepackage{makecell}
\usepackage{diagbox}
\usepackage{subfigure}
\usepackage{marvosym}
\usepackage{balance}

\begin{document}

\title{MTmixAtt: Integrating Mixture-of-Experts with Multi-Mix Attention for Large-Scale Recommendation}

\author{Xianyang Qi}
\authornotemark[1]
\author{Yuan Tian}
\authornote{Both authors contributed equally to this research.}
\affiliation{%
  \institution{Meituan}
  \city{Beijing}
  \country{China}
}

\email{qixianyang@meituan.com}
\email{tianyuan31@meituan.com}

\author{Zhaoyu Hu}
\authornote{Corresponding author.}


\affiliation{%
  \institution{Meituan}
  \city{Beijing}
  \country{China}
}
\email{huzhaoyu02@meituan.com}

\author{Zhirui Kuai}
\affiliation{%
  \institution{Meituan}
  \city{Beijing}
  \country{China}
}
\email{kuaizhirui@meituan.com}

\author{Chang Liu}

\affiliation{%
  \institution{Beijing University of Posts and Telecommunications}
  \city{Beijing}
  \country{China}
}
\email{changliu@bupt.edu.cn}

\author{Hongxiang Lin}

\affiliation{%
  \institution{Meituan}
  \city{Beijing}
  \country{China}
}
\email{linhongxiang02@meituan.com}

\author{Lei Wang}

\affiliation{%
  \institution{Meituan}
  \city{Beijing}
  \country{China}
}
\email{wanglei46@meituan.com}



\renewcommand{\shortauthors}{Hu et al.}

\begin{abstract}
Industrial recommender systems critically depend on high-quality ranking models. However, traditional pipelines still rely on manual feature engineering and scenario-specific architectures, 
which hinder cross-scenario transfer and large-scale deployment. To address these challenges, we propose \textbf{MTmixAtt}, a unified Mixture-of-Experts (MoE) architecture with Multi-Mix Attention, 
designed for large-scale recommendation tasks. MTmixAtt integrates two key components. 
The \textbf{AutoToken} module automatically clusters heterogeneous features into semantically coherent tokens, removing the need for human-defined feature groups. 
The \textbf{MTmixAttBlock} module enables efficient token interaction via a learnable mixing matrix, 
shared dense experts, and scenario-aware sparse experts, capturing both global patterns and scenario-specific behaviors within a single framework. Extensive experiments on the industrial TRec dataset from Meituan demonstrate that MTmixAtt consistently outperforms state-of-the-art baselines including Transformer-based models, WuKong, HiFormer, MLP-Mixer, and RankMixer. 
At comparable parameter scales, MTmixAtt achieves superior CTR and CTCVR metrics; scaling to MTmixAtt-1B yields further monotonic gains. 
Large-scale online A/B tests validate the real-world impact: 
in the \textit{Homepage} scenario, MTmixAtt increases Payment PV by \textbf{+3.62\%} 
and Actual Payment GTV by \textbf{+2.54\%}. Overall, MTmixAtt provides a unified and scalable solution for modeling arbitrary heterogeneous features across scenarios, 
significantly improving both user experience and commercial outcomes.
\end{abstract}




\maketitle

\begin{figure*}[h]
\centering
\includegraphics[width=2\columnwidth]{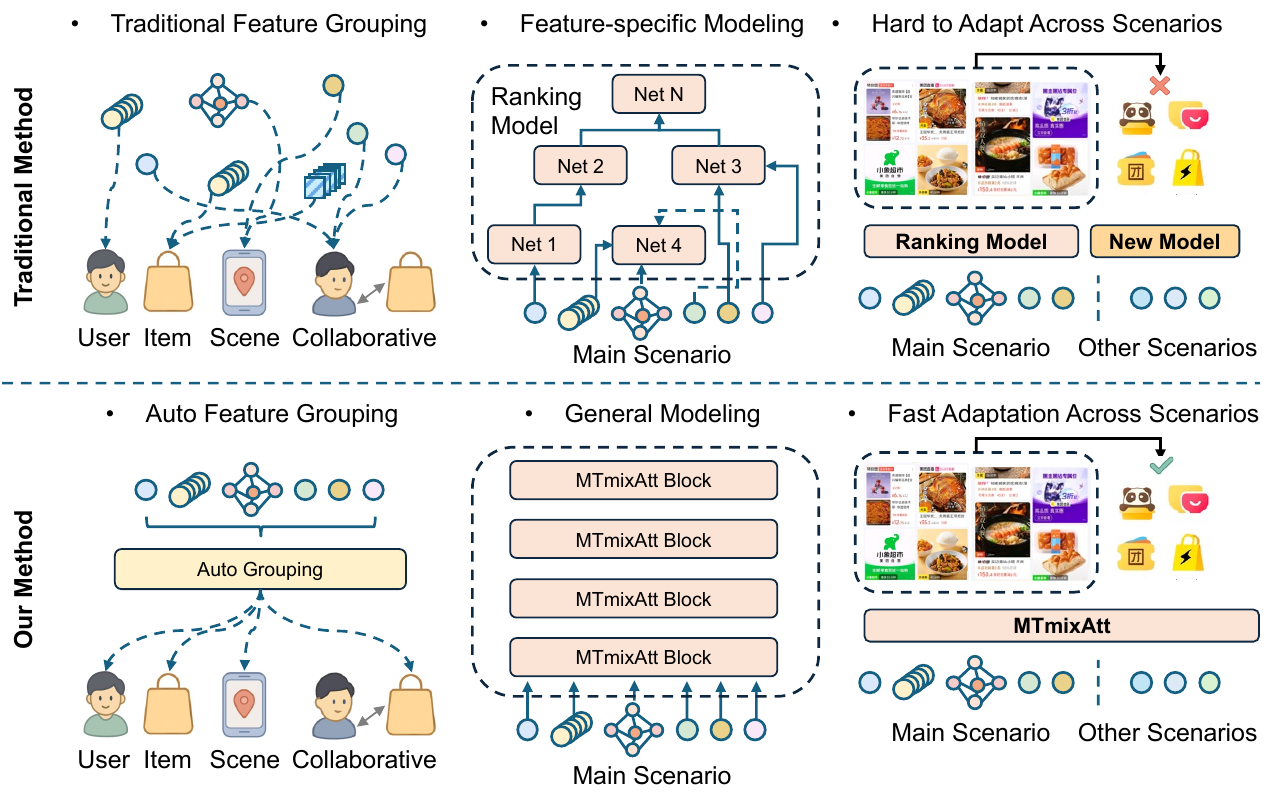}
\caption{Typical ranking models (top) rely on manual feature grouping and scenario-specific networks, leading to poor cross-scenario transferability. MTmixAtt (bottom) introduces automatic grouping and a unified architecture, enabling arbitrary feature modeling and fast cross-scenario adaptation.}
\label{fig}
\end{figure*}

\section{Introduction}

Recommender systems form the backbone of modern online platforms such as e-commerce, short-video, and local-life applications. 
Among them, ranking models play a pivotal role in delivering personalized and accurate recommendations by estimating the relevance between users and candidate items~\cite{Ref1, ff1, ff2, ff3}. 
The performance of large-scale recommender systems thus critically depends on the design of effective ranking architectures~\cite{Ref2, ff5}.

Despite their success, existing industrial ranking models still face fundamental challenges. 
\textbf{First}, feature engineering remains the primary bottleneck. 
Features are typically constructed and grouped under scenario-specific assumptions, where practitioners manually cluster heterogeneous signals (e.g., user profiles, item attributes, and interaction histories) into semantically coherent groups~\cite{borisyuk2024lirankindustriallargescale}. 
This manual process is highly subjective and inconsistent across practitioners, often leading to performance variance even under the same scenario. 
\textbf{Second}, architectural heterogeneity severely limits scalability. 
Different feature modalities require different network structures—dense features rely on MLPs, sequential features on RNNs or Transformers, and graph-structured data on GNNs~\cite{GNNs}. 
Such non-uniform designs complicate system integration and prevent straightforward parameter scaling. 
\textbf{Third}, cross-scenario generalization remains limited in multi-scenario industrial platforms such as Meituan~\cite{hu2025dynamicforgettingspatiotemporalperiodic}, Douyin~\cite{lan2025nextuserretrievalenhancingcoldstart}, and Kuaishou~\cite{wang2024homehierarchymultigateexperts}. 
A model optimized for one scenario (e.g., \textit{Homepage}) often fails to generalize to another (e.g., \textit{Groupon Promotions}), requiring costly re-engineering~\cite{multicene}.

Recent works such as RankMixer~\cite{zhu2025rankmixerscalingrankingmodels}, inspired by the scaling laws in large language models, 
have explored unified architectures for efficient scaling and heterogeneous feature modeling. 
However, RankMixer still inherits the limitations above—it relies on manual feature grouping and lacks cross-scenario adaptability. 
Our empirical analysis further shows that grouping strategies substantially affect model performance, yet current approaches remain dependent on heuristic, domain-specific decisions.

To overcome these challenges, we propose \textbf{MTmixAtt}, a \textbf{M}ul\textbf{T}i-\textbf{mix} \textbf{Att}ention Mixture-of-Experts (MoE) architecture for recommendation ranking. 
Here, \textit{Multi} reflects the multi-head and multi-scenario nature of the architecture, 
while \textit{Mix} denotes the unified design that integrates token mixing and expert mixing within an attention-like formulation. 
MTmixAtt unifies feature grouping, heterogeneous feature modeling, and multi-scenario adaptation within a single scalable framework. Specifically, it enables: 
(1) automatic feature grouping without human intervention, 
(2) unified modeling of heterogeneous features through a token-mixing MoE backbone, 
and (3) efficient multi-scenario adaptation with shared and scenario-specific experts. 
We deploy MTmixAtt in a large-scale local-life recommendation system serving hundreds of millions of daily active users, 
scaling up to one billion parameters and achieving consistent improvements over strong industrial baselines.

\textbf{Our contributions are threefold:}
\begin{itemize}[leftmargin=10pt]
\item We propose \textbf{MTmixAtt}, a unified token-mixing MoE architecture for large-scale recommendation ranking that jointly handles automatic feature grouping, heterogeneous feature modeling, and multi-scenario adaptation.
\item We design an auto-grouping mechanism that discovers feature clusters in a data-driven manner and an expert routing strategy that integrates shared and scenario-specific experts for efficient knowledge transfer.
\item We validate MTmixAtt through extensive offline experiments on the large-scale TRec dataset and online A/B tests in Meituan’s production environment, demonstrating consistent improvements over state-of-the-art baselines and significant business gains.
\end{itemize}

\section{Related Work}
\subsection{Ranking Models in Recommender Systems}

Before the deep learning era, ranking models in recommender systems primarily relied on manual feature engineering~\cite{Su_2022, bu1}. 
Logistic regression with handcrafted cross features was the dominant approach, 
while tree-based methods such as Gradient Boosted Decision Trees (GBDT)~\cite{lin2025onlinegradientboostingdecision} 
were also used to capture limited nonlinear feature interactions. 
The introduction of Factorization Machines (FMs)~\cite{FM} 
revolutionized this paradigm by enabling efficient second-order feature interaction modeling through latent vector inner products. 
With the rise of deep learning, Wide \& Deep~\cite{cheng2016widedeeplearning} combined explicit cross features with implicit DNN-based representation learning, 
inspiring a range of hybrid architectures such as DeepFM~\cite{guo2018deepfmendtoendwide}, 
PNN~\cite{PNN}, and DCN/DCN-V2~\cite{DCNv2}. 
More recently, attention-based architectures such as AutoInt~\cite{Autoint} 
have been proposed to model higher-order and dynamic feature interactions. 
Despite these advances, most models remain tailored to specific feature structures and lack the universality required for complex, multi-scenario recommendation tasks.

\subsection{Modeling Heterogeneous Features and Scaling}

While ranking models focus on feature interaction design, 
modern recommender systems must also address the challenge of handling heterogeneous inputs, 
including dense numerical signals, sparse categorical embeddings, sequential behaviors, and multimodal content. 
Different modules have been developed to process these feature types: 
MLPs for dense features, Transformers for behavioral sequences, 
GNNs for item-to-item relations, and multimodal encoders for text or image content. 
However, such designs often rely on task-specific modules, resulting in fragmented architectures that are difficult to scale or unify.

To improve generalization and scalability, several works have proposed unified architectures. 
Hiformer~\cite{hiformer} leverages heterogeneous attention layers to process different feature types in a shared framework, 
while WuKong~\cite{wukong} employs stacked factorization machines combined with a synergistic upscaling strategy for large-scale recommendation. 
RankMixer~\cite{zhu2025rankmixerscalingrankingmodels} continues this direction with a hardware-efficient token mixing mechanism 
and sparse mixture-of-experts (MoE) design for billion-scale deployment. 
Despite these advancements, current approaches still face notable challenges: 
they lack the flexibility to adapt to diverse feature spaces and scenario heterogeneity.

In summary, existing research has made significant progress in ranking optimization, feature interaction design, and heterogeneous modeling, 
but remains fragmented and domain-specific. 
A unified, general-purpose ranking framework that jointly handles feature grouping, heterogeneity, and scalability across scenarios is still an open problem.

\section{Method}
\subsection{Overview}
In recommendation systems, all features are referred to as $C$. The ranking task can be formally expressed through our proposed model architecture as:
\begin{equation}
    \hat{y} = \text{MTmixAtt}(G(C)),
\end{equation}
where $G(C)$ denotes an automatic grouping function responsible for adaptively dividing the features $C$ which include encoded sequence features and other features (multimodal features, user attributes, item attributes, user-item cross features, and scenario-specific features) into semantically coherent groups to facilitate enhanced feature interactions. $\text{MTmixAtt}(\cdot)$ is a mixed attention module that enables intra-scenario and inter-scenario feature interactions, and $\hat{y}$ is the final prediction output.

The two core components of this architecture operate in close synergy, and each is detailed in subsequent sections. Specifically, the automatic grouping module is referred to as \textbf{AutoToken} and elaborated on in \textit{Section 3.2}. The mixed attention module is defined as \textbf{MTmixAttBlock} and discussed in \textit{Section 3.3}.

\subsection{AutoToken}\label{AA}
In feature interaction modeling for recommendation systems, the rationality of feature clustering strategies directly influences the model's ability to capture semantic relationships and its predictive performance. Theoretically, improper clustering may introduce noise and disrupt intrinsic semantic connections between features, thereby reducing the model's expressive power and even leading to performance instability and degraded generalization. Traditional approaches often rely on manually designed clustering rules, which exhibit significant limitations: on the one hand, human-designed rules are highly subjective and lack guarantees of objectivity and consistency; on the other hand, they fail to flexibly adapt to varying data distributions and scenario requirements, severely restricting the portability and optimization of the model.

\begin{figure}[h]
\centering
\includegraphics[width=1\columnwidth]{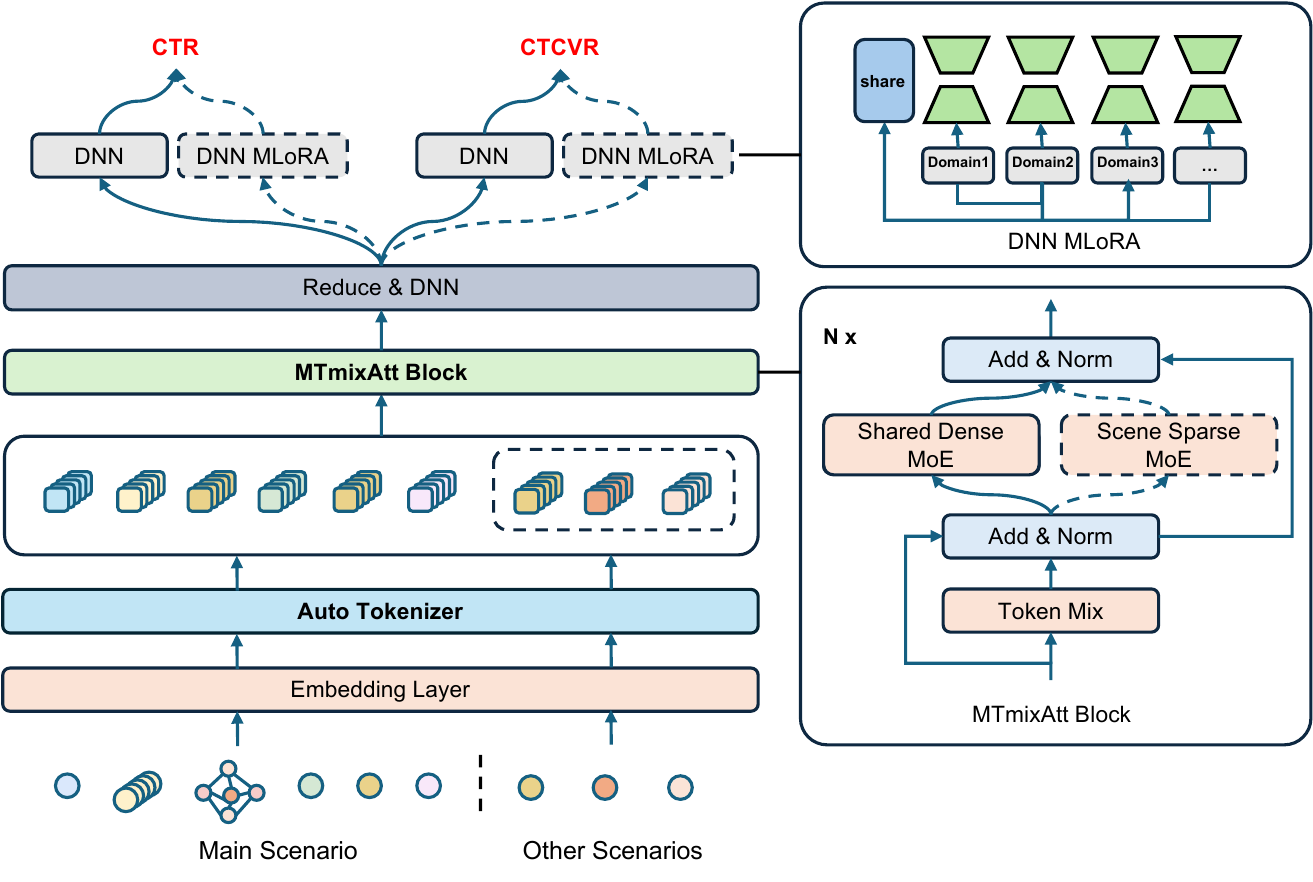}
\caption{The architecture of MTmixAtt.}
\label{fig2}
\end{figure}

To address the limitations of manual feature grouping, we propose \textbf{AutoToken}, a data-driven mechanism that automatically learns to organize heterogeneous features into groups. This approach mitigates human bias and enhances cross-scenario adaptability by allowing the model to infer grouping structures directly from data.

The concrete implementation involves the following steps. First, dimension alignment is performed on the raw features via nonlinear transformation:
\begin{equation}
\hat{x}_i = \text{DNN}_i(x_i), \quad i = 1,2,\ldots,n_f,
\end{equation}
where $x_i$ denotes the $i$-th original feature, $\hat{x}_i$ is its dimension-aligned representation, $\text{DNN}_i$ refers to the multi-layer perceptron assigned to the $i$-th feature, and $n_f$ is the total number of features. The aligned features form the matrix:
\begin{equation}
\hat{X} = [\hat{x}_1, \hat{x}_2, \ldots, \hat{x}_{n_f}] \in \mathbb{R}^{n_f \times e},
\end{equation}
where $e$ denotes the unified embedding dimension.

Then, to achieve dynamic grouping, a learnable feature selection matrix 
$W \in \mathbb{R}^{n_g \times n_f}$ 
is constructed, where $n_g$ denotes the number of groups. 
During forward computation, grouped representations are obtained via 
Top-$k$ index selection and weighted aggregation:
\begin{equation}
X = G(T_k(W), \hat{X}) \cdot S_k(W),
\label{eq:grouping}
\end{equation}
where $T_k(W_i) = \arg\max\text{-}k(W_i)$ selects the 
top-$k$ indices from the $i$-th row of $W$, and 
$G(I, \hat{X}) = \mathrm{concat}(\hat{x}_j),\, j \in I$ 
aggregates the feature vectors corresponding to the index set $I$. 
Here, $I$ is the set of selected indices, and 
$\hat{\mathbf{x}}_j$ is the $j$-th aligned feature vector. 
The grouped representation $G(I, \hat{\mathbf{X}})$ has the shape 
$\mathbb{R}^{k \times e \times n_g}$, which is reshaped to 
$\mathbb{R}^{d \times n_g}$ by flattening the first two dimensions 
($d = k \times e$).

To ensure gradient backpropagation through $W$, 
the weights are normalized using:
\begin{equation}
S_k(W_i) = \mathrm{softmax}(\mathrm{top}\text{-}k(W_i)),
\label{eq:softmax}
\end{equation}
where $S_k(W_i) \in \mathbb{R}^{k}$ outputs the 
softmax-normalized scores of the top-$k$ values in $W_i$. 
This operation makes the entire clustering process differentiable and optimizable.

\subsection{MTmixAttBlock}

Inspired by the RankMixer architecture~\cite{zhu2025rankmixerscalingrankingmodels}, we propose a redesigned feature interaction architecture termed MTmixAttBlock. As illustrated in Figure~\ref{fig2}, the module integrates three core components: 
(1) a learnable token mixing module that enables fine-grained feature interactions across semantic subspaces; 
(2) a shared dense MoE module that enhances the diversity of learned representations; and 
(3) a multi-scenario expert network that allows adaptive modeling across different contexts. 
Unlike RankMixer, which applies static token mixing, our design introduces head-wise learnable mixing matrices that dynamically capture complex dependencies among feature groups. Conceptually, token mixing plays a role analogous to the tokenizer, while the gating mechanism in dense MoE models parallels the attention weighting process, hence the name \textbf{MTmixAttBlock}.

\subsubsection{Learnable Mixing Matrix}

To strengthen feature interactions across token dimensions, we introduce a \textit{learnable mixing matrix} mechanism. 
Before mixing, the input feature matrix is transposed such that the token dimension becomes the primary axis of transformation. 
Formally, given the aligned feature representation $X \in \mathbb{R}^{d \times n_g}$, where $d$ denotes the feature dimension and $n_g$ represents the Top-$k$ tokens selected by AutoToken, we first apply a transposition to enable token-wise processing:
\begin{equation}
M = X^{\top} \in \mathbb{R}^{n_g \times d}.
\end{equation}

The transposed matrix $M$ is then divided into $H$ attention heads along the feature dimension:
\begin{equation}
M = \big[\,M^{(1)} \;\|\; M^{(2)} \;\|\; \cdots \;\|\; M^{(H)} \big], 
\quad M^{(h)} \in \mathbb{R}^{n_g \times d_h}, \; d_h=d/H.
\end{equation}

For each head $h$ in $H$, a learnable transformation matrix $W_h \in \mathbb{R}^{n_g \times n_g}$ 
is applied along the token dimension to perform head-wise feature interaction:
\begin{equation}
Y^{(h)} = W_h^{\top} M^{(h)}.
\end{equation}
Here, $W_h$ determines how information flows among tokens within each semantic head.

A residual connection is employed to stabilize optimization and preserve the original semantics:
\begin{equation}
Z^{(h)} = M^{(h)} + Y^{(h)} = \left(I + W_h^{\top}\right) M^{(h)}.
\end{equation}

Finally, the outputs from all heads are concatenated along the feature dimension to form the final representation:
\begin{equation}
Z = \big[\,Z^{(1)} \;\|\; Z^{(2)} \;\|\; \cdots \;\|\; Z^{(H)} \big] \in \mathbb{R}^{n_g \times d}.
\end{equation}

This design ensures that token-level interactions are explicitly modeled after transposition, 
allowing the model to flexibly capture dependencies across token positions while maintaining computational efficiency.

\begin{figure}[h]
\centering
\includegraphics[width=1\columnwidth]{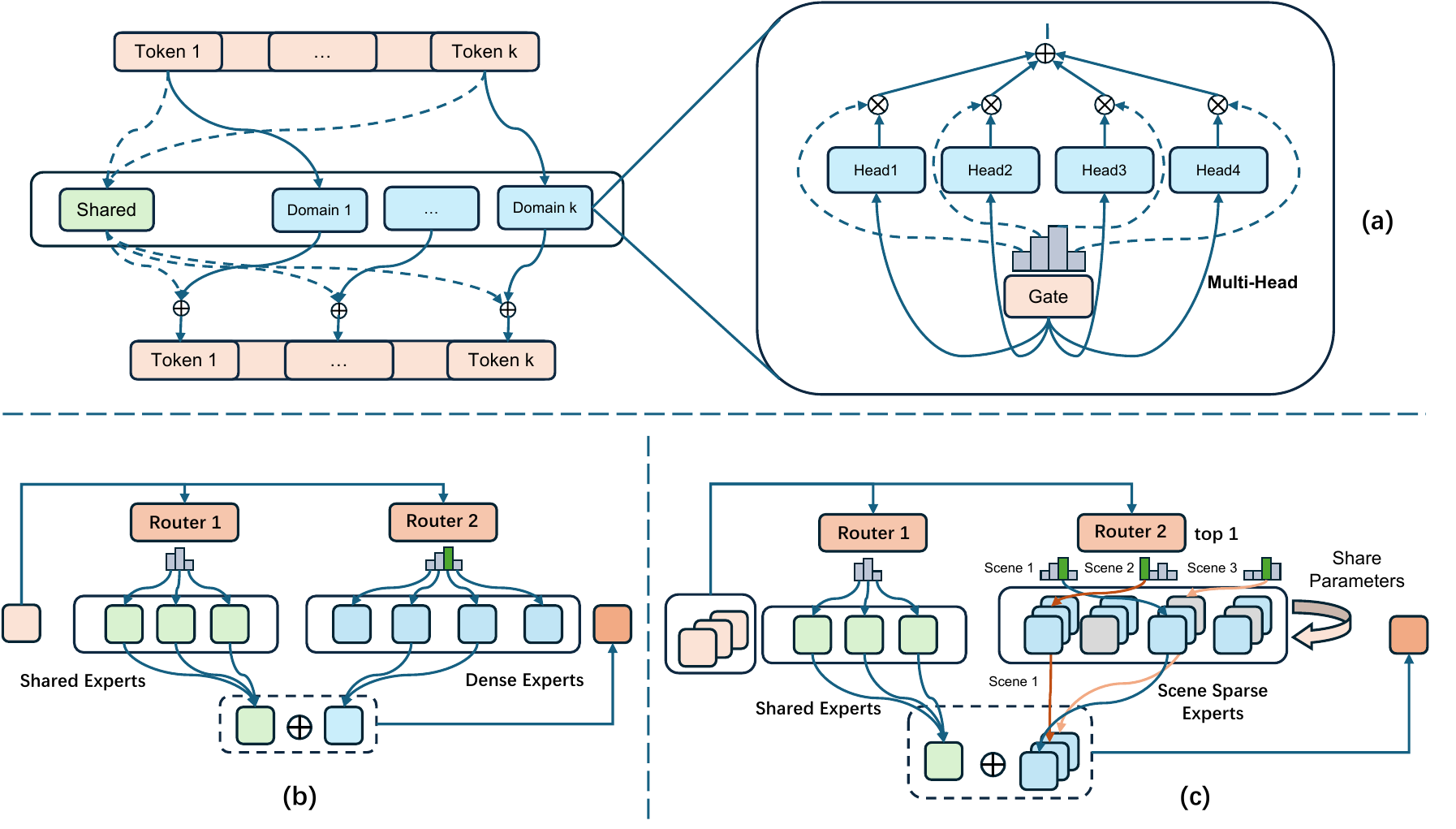}
\caption{Illustration of the Shared Dense MoE and Scenario-Specific Sparse MoE. (a) Overall MoE architecture. (b) Shared Dense MoE under the main scenario. (c) Scenario-Specific Sparse MoE across multiple scenarios.}
\label{fig3}
\end{figure}

\subsubsection{Shared Dense MoE}
\label{Shared_Dense_MoE}

After token-level interaction is performed by the learnable mixing matrix, 
we further introduce a \textit{Shared Dense MoE} layer to enhance semantic specialization across tokens. 
This component allows the model to allocate different experts to different token patterns, 
thereby improving representation diversity and cross-token semantic integration.

RankMixer~\cite{zhu2025rankmixerscalingrankingmodels} employs token grouping for disentangled modeling of feature groups, 
but such groups are not entirely independent in practice (e.g., user click and payment histories are correlated). 
To better capture these inter-group relationships, we refine the expert allocation strategy following 
recent advances in mixture-of-experts design~\cite{dai2024deepseekmoeultimateexpertspecialization}. 
As illustrated in Figure~\ref{fig3}, the proposed Shared Dense MoE consists of two components.

\paragraph{Fine-grained Expert Splitting.}
Previous studies~\cite{dai2024deepseekmoeultimateexpertspecialization} have shown that increasing model size alone often leads to diminishing returns, 
as overly large experts may fail to capture diverse perspectives. 
To address this, we adopt a fine-grained expert strategy by splitting each feed-forward expert (FFN) into $m$ smaller sub-experts, 
thus expanding the number of available experts while keeping the total computational cost fixed. 
This finer granularity enables more flexible expert combinations and enhances the model's adaptability.

\paragraph{Shared Expert Isolation.}
Stable global features in recommendation systems (e.g., long-term user profiles) 
often serve as shared semantic anchors~\cite{chang2023pepnet, chai2025longer, wang2025home}. 
If each group of features is modeled independently, experts may redundantly learn these anchors or fail to reuse shared information. 
To address this issue, we introduce a small set of shared experts that are always activated for every token, 
capturing generalizable knowledge across all feature groups. 
This shared mechanism improves parameter efficiency and reduces redundancy while maintaining model expressiveness.

Formally, the token-level expert aggregation at the $l$-th layer is defined as:
\begin{equation}
\label{eq:shared_moe}
h_t^l = 
\underbrace{\sum_{i=1}^{K_s} \alpha_{i,t} \, \text{FFN}_i(u_t^l)}_{\text{Shared experts}} 
+ 
\underbrace{\sum_{j=1}^{mN} \beta_{j,t} \, \text{FFN}_j(u_t^l)}_{\text{Fine-grained experts}},
\end{equation}
where $u_t^l \in \mathbb{R}^{d}$ is the input representation of the $t$-th token at layer $l$ and $h_t^l \in \mathbb{R}^{d}$ is the aggregated output. 
$\text{FFN}_i(\cdot)$ denotes the feed-forward expert network assigned to expert $i$. 

Each expert is weighted by a gating function:
\begin{equation}
\alpha_{i,t} = \sigma(\text{Gate}_{1}(u_t^l)), \quad
\beta_{j,t} = \sigma(\text{Gate}_{2}(u_t^l)),
\end{equation}
where $\sigma(\cdot)$ is the sigmoid activation that produces weights in $[0,1]$. 
The first term represents the $K_s$ shared experts that are always active, 
while the second term corresponds to $mN$ fine-grained experts 
obtained by splitting each of the $N$ base experts into $m$ smaller sub-experts. 
These sub-experts are selectively activated according to the gating function.

\subsubsection{Multi-scenario Expert Networks}

In multi-scenario modeling, traditional methods either use fully shared parameters, which lack scenario-specific adaptation, or build independent models for each scenario, which leads to fragmented knowledge and redundant parameters \cite{chang2023pepnet, zhou2023hinet, tian2023multi}. To address these issues, we propose a sparse hybrid architecture, which integrates both shared experts and scenario-specific experts at the scenario level. This architecture enables a unified modeling framework: shared experts learn universal interaction patterns across scenarios, while scenario experts capture behaviors unique to each scenario.

\paragraph{Shared Experts Across Multi-Scenario.}
As illustrated in Figure~\ref{fig3}(c), in addition to the token-level shared experts (the left term in Equation~11), 
the fine-grained expert layer further introduces scenario-level shared experts. 
In this setting, multiple samples or scenarios are input while the model outputs a single token representation. 
The same token shares experts across different scenarios, 
which effectively prevents interference between general and scenario-specific knowledge when both are learned within the same expert. 
Let $u_t^l \in \mathbb{R}^d$ denote the representation of token $t$ at layer $l$, 
and let $\mathbf u_c^l \in \mathbb{R}^{d}$ denote the representation of scenario token $c$ at layer l. 
The corresponding gating weight for shared experts is computed as:
\begin{equation}
p_{i,t} = \sigma(\text{Gate}_{3}(u_t^l\|\, u_c^l)),
\end{equation}

\paragraph{Sparse Expert Selection Across Multi-scenario.}
In multi-scenario modeling, one approach is to build separate models for each scenario with isolated parameters, but this greatly increases the number of model parameters. 
To balance effectiveness and efficiency, we introduce sparse scenario experts across different scenarios to reduce conflicts caused by heterogeneous feature spaces. 
Each sample is processed by both shared experts and scenario-specific experts for every token. 
Shared experts process all samples, while scenario-specific experts are activated with weights determined by the scenario. 
Additionally, \text{Top-}k other experts are selected to encourage sharing while maintaining sparse isolation. We first compute the gating logits as
\begin{equation}
z_{i,t} = 
\mathrm{Gate}_4\!\left(\left[\,u_t^l \,\|\, u_c^l\,\right]\right)
+ \gamma\,\mathbbm{1}[\,i = i^*\,],
\end{equation}
where $\gamma>0$ is a manually assigned bonus that enforces activation of the current scenario expert $i^*$. We then perform sparse routing by applying a mask $m_{i,t}$ that retains only the Top-$k$ experts with the highest routing scores for each token, effectively filtering out low-scoring experts and ensuring efficient expert utilization. Finally, the masked routing logits are
\begin{equation}
\tilde{z}_{i,t} = z_{i,t} \cdot m_{i,t}.
\end{equation}

Next, we apply a sigmoid activation to obtain the normalized routing scores:
\begin{equation}
s_{i,t} = \sigma(\tilde{z}_{i,t}),
\end{equation}
where $\sigma(\cdot)$ denotes the sigmoid function that produces values in $(0,1)$. 
The scenario-level sparse gating is then formulated as:

\begin{equation}
\beta_{i,t} =
\begin{cases}
s_{i,t} + p_{i,t}, & s_{i,t} \in \text{Top-}k(s_{j,t}, mK), \\[6pt]
0, & \text{otherwise}.
\end{cases}
\end{equation}

$\text{Top-}k(\cdot, mK)$ returns the set of $mK$ experts with the highest affinity scores for the $t$-th token among all candidates. The computation for the shared experts on the left-hand side of Eq.~(11) remains unchanged,
while the \textit{Fine-grained experts} term on the right-hand side is replaced by the sparse gating weights $\beta_{i,t}$ from Eq.~(17),
yielding the output of the $l$-th layer.

\subsubsection{Output Layer and Optimization Objective.}

The aggregated token representations from the expert layers are passed to scenario-specific prediction heads to produce the final outputs. 
For the \emph{main scenario}, a standard feed-forward DNN module is employed to predict the target score $\hat{y}^{(\text{m})}$. 
For \emph{other scenarios}, we attach lightweight adapters based on MLoRA~\cite{yang2024mloramultidomainlowrankadaptive} 
to the same backbone network, enabling efficient parameter sharing across scenarios while retaining domain-specific adaptability. 
Each adapter introduces a small set of low-rank trainable parameters that specialize the shared DNN for its corresponding scenario.

Formally, let $y^{(c)}$ and $\hat{y}^{(c)}$ denote the ground-truth label and predicted probability 
for a sample from scenario $c \in \{1, \ldots, C\}$. 
The objective function minimizes the average binary cross-entropy loss across all scenarios:
\begin{equation}
\label{eq:loss}
\mathcal{L} 
= - \frac{1}{C} \sum_{c=1}^{C} 
    \mathbb{E}_{(x,y)\sim \mathcal{D}_c} 
    \Big[ 
        y^{(c)} \log \hat{y}^{(c)} 
        + (1 - y^{(c)}) \log (1 - \hat{y}^{(c)}) 
    \Big],
\end{equation}
where $\mathcal{D}_c$ denotes the data distribution of scenario $c$. 
This formulation jointly optimizes the shared backbone and scenario-specific MLoRA adapters in an end-to-end manner, 
achieving a balance between global knowledge sharing and local adaptation.

\section{Experiments}
\subsection{Experimental Setting}
\begin{table}[ht]
\caption{Performance comparison across different models in terms of CTR/CTCVR AUC and GAUC. MTmixAtt-1B achieves the best performance across all metrics.}
    \centering
    \begin{adjustbox}{max width=0.47\textwidth}
    \begin{tabular}{cccccc}
    \toprule
    \textbf{Models} & \makecell{\textbf{CTR}\\\textbf{AUC}} & \makecell{\textbf{CTR}\\\textbf{GAUC}} & \makecell{\textbf{CTCVR}\\\textbf{AUC}} & \makecell{\textbf{CTCVR}\\\textbf{GAUC}} & \textbf{Params} \\
    \midrule
    Online Base        & 0.7771 & 0.6915 & 0.8828 & 0.7649 & 11M \\
    \midrule
    Transformer & 0.7722 & 0.6860 & 0.8791 & 0.7592 & 28M \\
    Wukong              & 0.7733 & 0.6870 & 0.8799 & 0.7611 & 21M \\
    Hiformer            & 0.7775 & 0.6909 & 0.8825 & 0.7662 & 15M \\
    MLP-Mixer           & 0.7713 & 0.6852 & 0.8785 & 0.7570 & 18M \\
    RankMixer           & 0.7781 & 0.6920 & 0.8832 & 0.7658 & 14M \\
    \midrule
    MTmixAtt-15M             & 0.7792 & 0.6931 & 0.8840 & 0.7682 & 15M \\
    MTmixAtt-1B              & \textbf{0.7811} & \textbf{0.6963} & \textbf{0.8858} & \textbf{0.7714} & \textbf{1B} \\
    \bottomrule
    \end{tabular}
    \end{adjustbox}
    \label{tab:overall3}
\end{table}

\textbf{Dataset.} The TRec dataset is constructed from large-scale, industrial purchase logs collected on Meituan, a leading e-commerce platform. It contains behavioral data spanning hundreds of millions of users over a two-month period. Following standard practice, we perform a time-based split: records from days [1, T] are used for model training, while data from day T+1 serves as the test set. TRec encompasses 786 million users, 162 million items, and 413 features, offering a rich feature space organized into six categories: sequential features, multimodal features, user attributes, item attributes, user-item cross features, and scenario-specific features.

\textbf{Evaluation Metrics.} We adapt Area Under the ROC Curve (AUC) and user Group AUC (G-AUC) as the primary evaluation metrics for both Click-Through Rate (CTR) and Click-to-Conversion Rate (CTCVR) tasks, which are widely used in both academia and industry. AUC measures the probability that the model ranks a randomly chosen positive instance higher than a randomly chosen negative one, reflecting the overall discriminative ability of the ranking model. G-AUC further evaluates model performance at the group level, where users or scenarios are treated as independent groups, and the final score is computed as a weighted average over groups.
\begin{figure}[h]
\centering
\includegraphics[width=1\columnwidth]{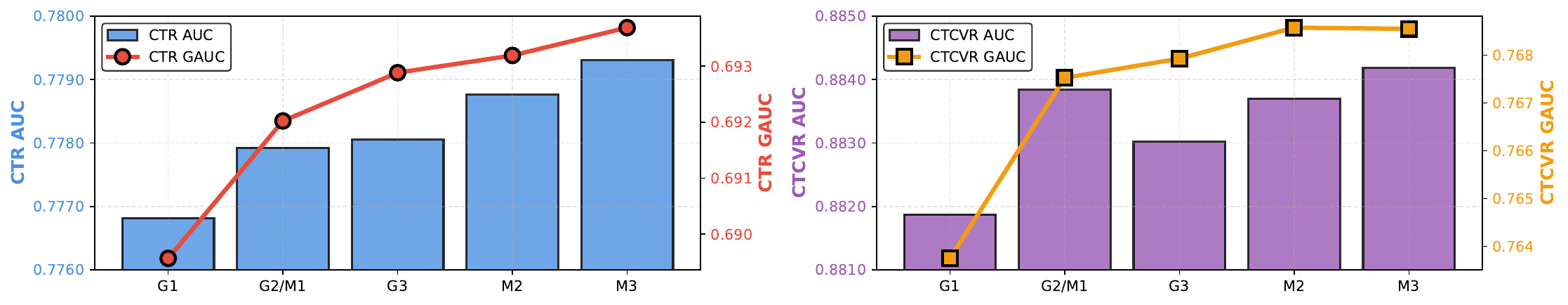}
\caption{Effects of AutoToken and the transformation matrix in token mixing module.}
\label{fig4}
\end{figure}

\begin{figure}[h]
\centering
\includegraphics[width=1\columnwidth]{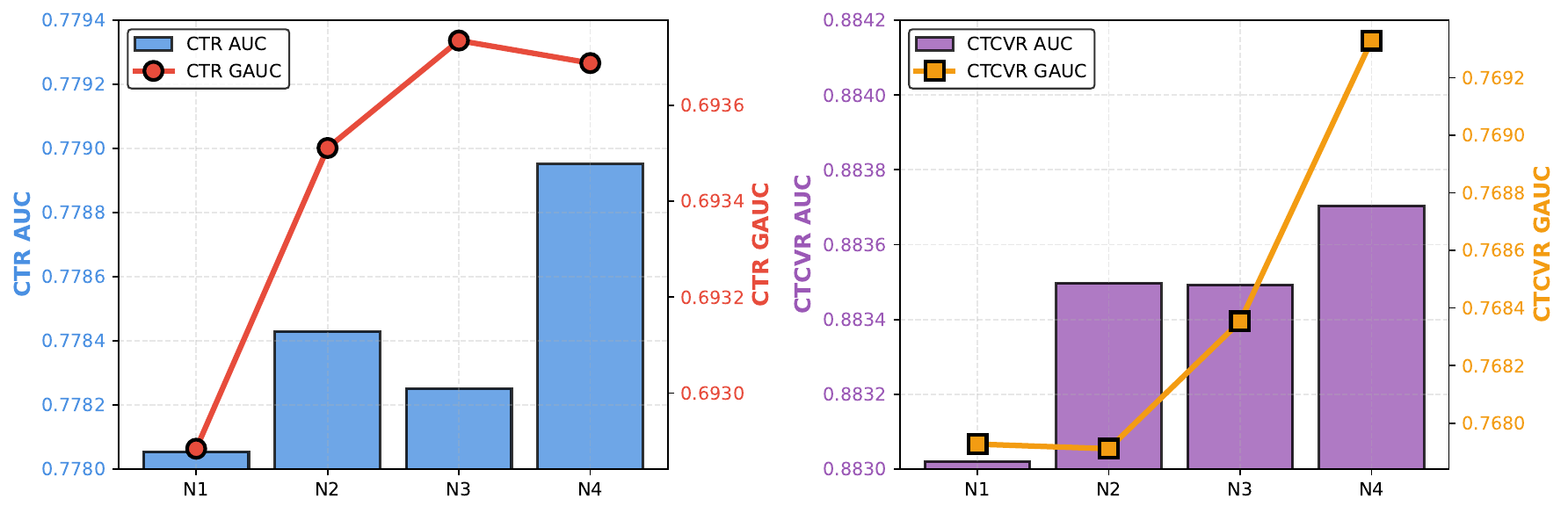}
\caption{Ablation experiments on dense MoE in the Homepage scenario.}
\label{fig5}
\end{figure}

\begin{figure*}[h]
\centering
\includegraphics[width=2\columnwidth]{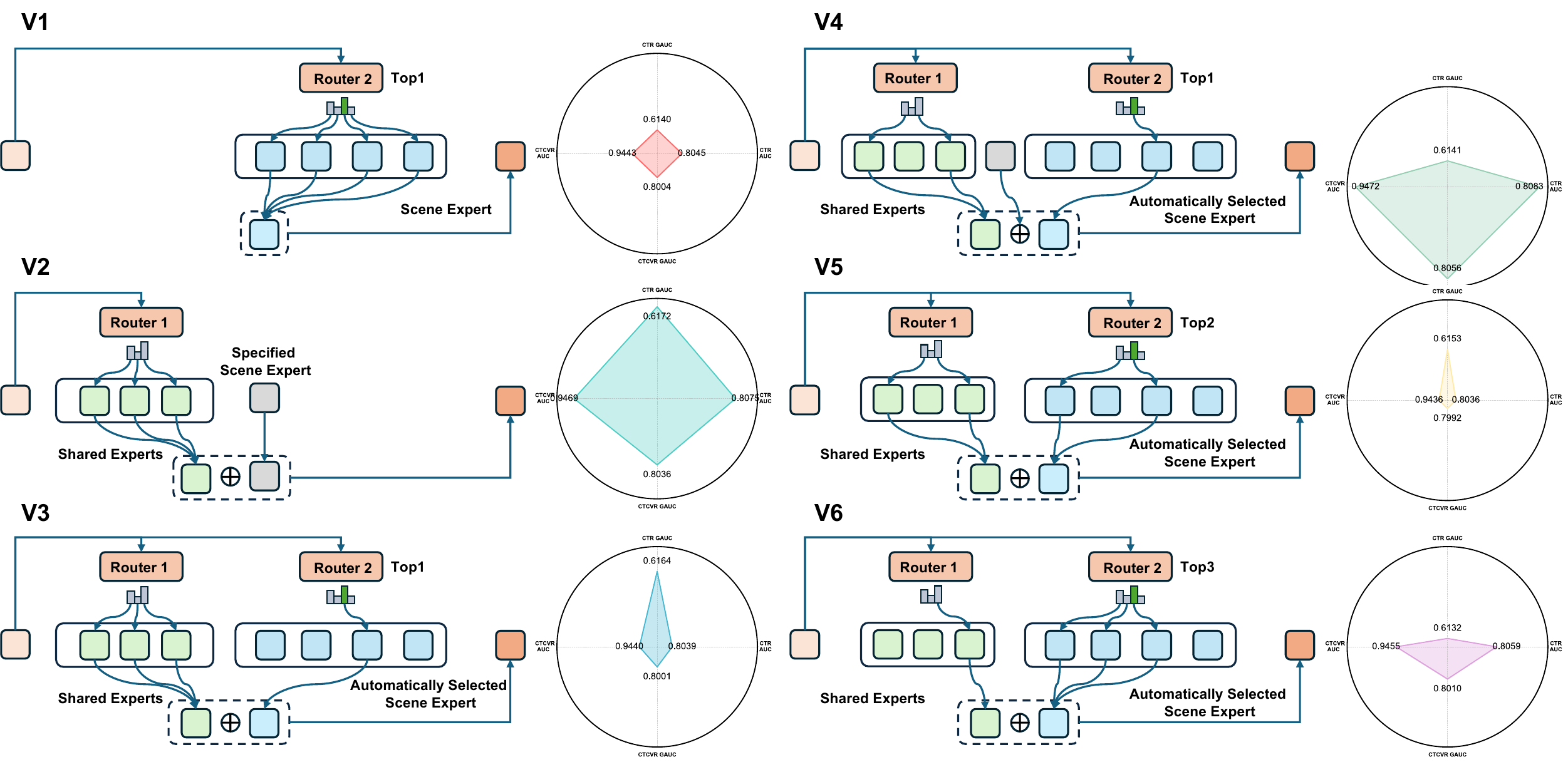}
\caption{Ablation experiments on scenario MoE in other scenarios.}
\label{fig6}
\end{figure*}

\textbf{Baselines.} To rigorously evaluate the effectiveness of MTmixAtt, we compare it against four categories of state-of-the-art methods. Online Base refers to the production model deployed in our platform, which processes different feature types through separate specialized modules and combines them in a hybrid manner (e.g., DNNs and Transformers). Transformer-based models adopt stacked self-attention layers as the feature interaction module. WuKong \cite{wukong} introduces a Factorization Machine block together with a Linear Compression block, achieving state-of-the-art performance in Meta’s large-scale recommendation scenarios. HiFormer \cite{hiformer} proposes a heterogeneous self-attention mechanism that explicitly accounts for the heterogeneity of feature interactions. Finally, MLP-Mixer \cite{MLP-Mixer} and RankMixer \cite{zhu2025rankmixerscalingrankingmodels} employ a token-mixing paradigm for feature interaction. These baselines cover the mainstream paradigms of ranking models, ensuring a comprehensive evaluation of MTmixAtt.

\textbf{Implementation Details.} The MTmixAtt-15M model is trained on 2 nodes with a total of 16 A100 GPUs. Its architecture is configured with 4 layers, 3 experts, 12 tokens, and a token dimension of 204. For the larger MTmixAtt-1B model, training is conducted on 4 nodes with a total of 32 GPUs. It has 8 layers, 4 experts, 27 tokens, and a token dimension of 756. We adopt the Adam optimizer with a learning rate of $5 \times 10^{-5}$, and set the batch size to 4,800.

\subsection{Overall Performance}
The overall performance is summarized in Table \ref{tab:overall3}. At a comparable parameter scale, MTmixAtt-15M achieves consistent improvements over strong baselines. For instance, compared with HiFormer (15M) and RankMixer (14M), MTmixAtt-15M yields higher CTR AUC (+0.17\% and +0.11\%, respectively) and CTR GAUC (+0.32\% and +0.39\%, respectively). In terms of CTCVR, MTmixAtt-15M also surpasses these models, improving GAUC by up to 0.20\%. These results indicate that even at a lightweight scale, MTmixAtt is more effective in modeling heterogeneous features than existing architectures. At the billion-parameter scale, MTmixAtt-1B further pushes the performance boundary, achieving the best results across all metrics.
\subsection{Ablation Study}

\textbf{Ablation Study on AutoToken.} For token grouping, we compare three strategies: random grouping (\underline{G1}), fully data-driven AutoToken (\underline{G2}), and manually designed prior grouping (\underline{G3}), as shown in Figure 4. The results demonstrate that AutoToken achieves a substantial improvement over random grouping, highlighting its effectiveness in discovering meaningful feature clusters. Compared with manual prior grouping, AutoToken achieves higher CTCVR AUC while maintaining only marginal decreases in CTR AUC, CTR GAUC, and CTCVR GAUC. This suggests that AutoToken is able to approximate, and in some cases surpass, the grouping expertise accumulated through years of domain practice.

\textbf{Comparison of different transformation matrix initialization strategies in the token mixing module.} As shown in Figure \ref{fig4}, we compare three variants of the token mixing module: \underline{M1}, using a fixed transpose matrix; \underline{M2}, using a learnable transformation matrix initialized with zeros; and \underline{M3}, using a learnable transformation matrix initialized with an orthogonal matrix. The results show that when the transformation matrix is learnable (\underline{M2}), the model captures feature interactions more effectively than the fixed transpose setting (\underline{M1}), leading to substantial improvements across all four metrics. In contrast, initializing the matrix with all ones leads to rank-1 degeneration, where all token representations collapse into the same direction. More critically, such initialization distorts the initial state of residual connections, resulting in only limited gains. For \underline{M3}, orthogonal initialization provides a more favorable starting point: it reduces to the transpose transformation at initialization while allowing the model to further learn richer feature interactions, which translates into additional improvements on CTR metrics.

\textbf{Ablation experiments on dense MoE in the main scenario.} As shown in Figure \ref{fig5}, we conduct ablation studies on the dense MoE component under the main scenario, denoted as \underline{N1}–\underline{N4}. \underline{N1} adopts the ReLU-based Sparse MoE from RankMixer, while \underline{N2} replaces it with a Sigmoid-based Dense MoE. \underline{N3} introduces the proposed Shared Dense MoE, where the number of shared experts is increased while the number of unique experts is reduced, keeping the total number of experts unchanged. \underline{N4} further increases the number of shared experts to two. The results show that \underline{N2} substantially outperforms \underline{N1} across all four metrics, as dense activation enables more effective expert utilization. Compared with \underline{N2}, \underline{N3} maintains comparable performance despite reducing the number of unique experts, indicating that an excessive reliance on scenario-specific experts may lead to knowledge redundancy, while shared experts can promote cross-token knowledge interaction and improve parameter efficiency. Finally, \underline{N4} achieves additional gains, particularly on the CTCVR GAUC metric, suggesting that combining shared experts with token-specific experts facilitates knowledge transfer across tokens and leads to more efficient use of model capacity.

\textbf{Ablation experiments on scenario MoE in other scenarios.}
As shown in Figure \ref{fig6}, we further conducted ablations on the scene MoE under other scenarios, keeping the total number of activated experts fixed at four. 
Six variants (\underline{V1}--\underline{V6}) were tested: 
\underline{V1}, dense MoE with four experts; 
\underline{V2}, three shared experts plus one manually specified scenario-specific expert (by assigning a large weight); 
\underline{V3}, three shared experts plus one automatically selected scenario-specific expert; 
\underline{V4}, two shared experts, one manually specified scenario expert, and one automatically selected expert; 
\underline{V5}, two shared experts plus two automatically selected scenario experts; 
and \underline{V6}, one shared expert plus three automatically selected scenario experts. 
Among them, \underline{V2} and \underline{V4} achieved the best overall performance. 
Considering that CTCVR GAUC is the most critical metric in our business setting, we ultimately adopt \underline{V4}. 
These results indicate that \underline{V4} provides a balanced modeling framework that captures both general patterns and scenario-specific behaviors: 
shared experts specialize in learning cross-scenario interactions, while scenario experts focus on unique patterns of each scenario. 
This division of labor prevents interference between general and specific knowledge, thereby enhancing both model expressiveness and learning efficiency.
 
\begin{figure}[h]
\centering
\includegraphics[width=1\columnwidth]{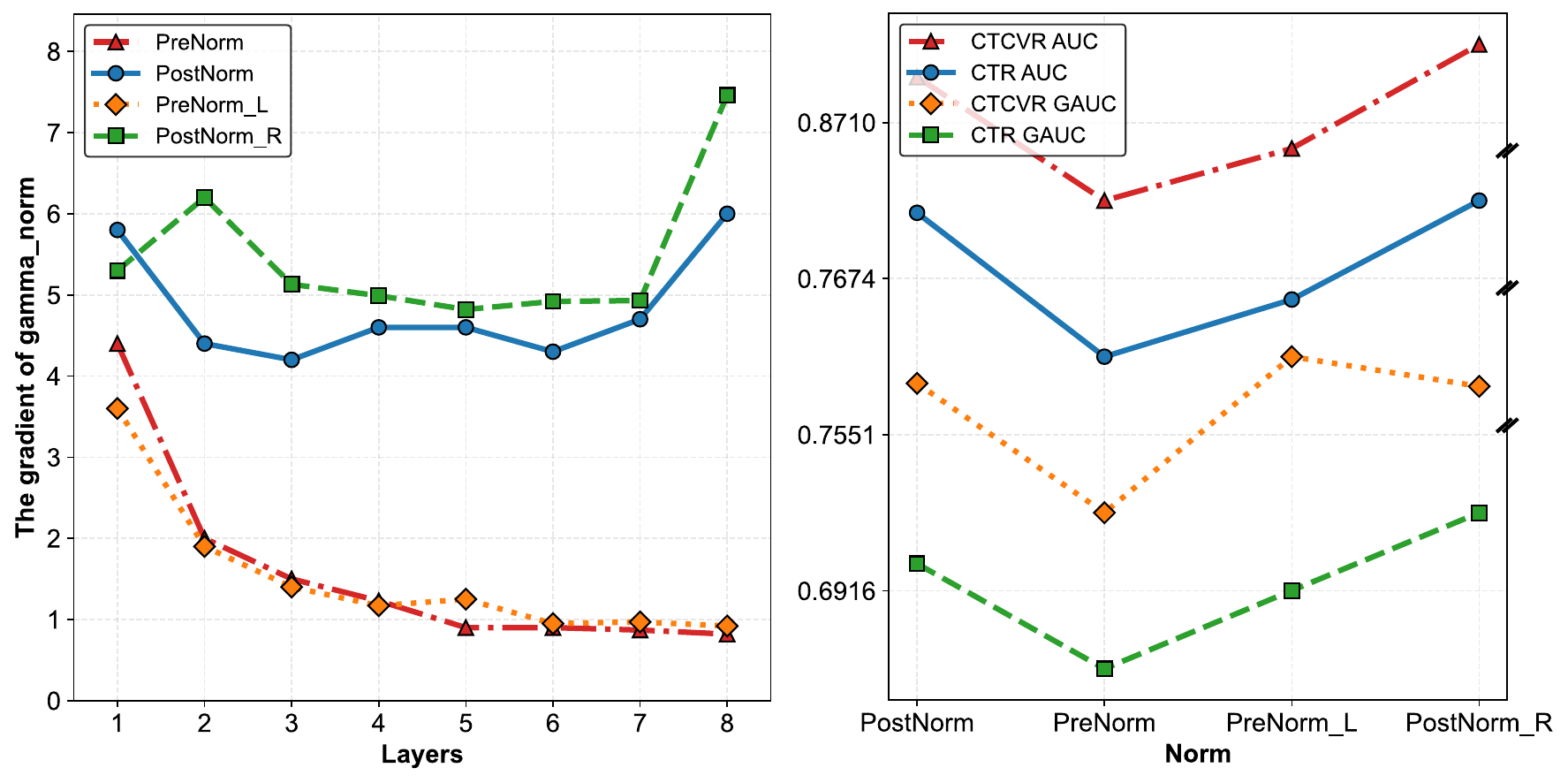}
\caption{Ablation experiments on Normalization. The left figure is a line chart showing gradient changes across different layers. The right figure presents the performance of four metrics for different normalization methods.}
\label{fig7}
\end{figure}

\textbf{Normalization experiments.} In large-scale language models (LLMs), normalization plays a critical role in ensuring training stability and facilitating information flow, especially in very deep networks \cite{deepseekai2025deepseekv3technicalreport, kimiteam2025kimik2openagentic}. To investigate whether similar trends exist in the MTmixAtt module, we examined four variants: PreNorm, PostNorm, PreNorm$_L$, and PostNorm$_R$. In PreNorm$_L$, an additional LayerNorm is applied at the final layer to mitigate gradient instability observed in standard PreNorm. In PostNorm$_R$, residual connections are added by summing the input to the output of each layer. Figure \ref{fig7} compares gradient dynamics across depths as well as performance on four evaluation metrics. Results show that PreNorm and PreNorm$_L$ suffer from noticeable gradient decay as the network depth increases, particularly in shallower layers. In contrast, both PostNorm and PostNorm$_R$ maintain stable gradients across all depths, with PostNorm$_R$ also achieving the best overall performance across metrics.

\begin{table}[ht]
\caption{Online lift of MTmixAtt in Homepage recommendation and other cross scenarios.}
    \centering
    \begin{adjustbox}{max width=\textwidth}
    \begin{tabular}{lccc}
\toprule
    \textbf{Scenario} & 
    \makecell{\textbf{Payment}\\\textbf{PV}} & 
    \makecell{\textbf{Actual Payment}\\\textbf{GTV}} & 
    \makecell{\textbf{NIEV}}\\
\midrule
Homepage          & +3.62\% & +2.54\% & +1.05\% \\
Cross Scene \#1   & +0.51\% & +1.02\% & +0.41\% \\
Cross Scene \#2   & +1.54\% & +2.71\% & +0.26\% \\
Cross Scene \#3   & +1.59\% & +1.54\% & +2.52\% \\
Cross Scene \#4   & +0.79\% & +1.29\% & +0.32\% \\
\bottomrule
    \end{tabular}
    \end{adjustbox}
    \label{tab:feed_metrics}
\end{table}

\begin{figure}[h]
\centering
\includegraphics[width=1\columnwidth]{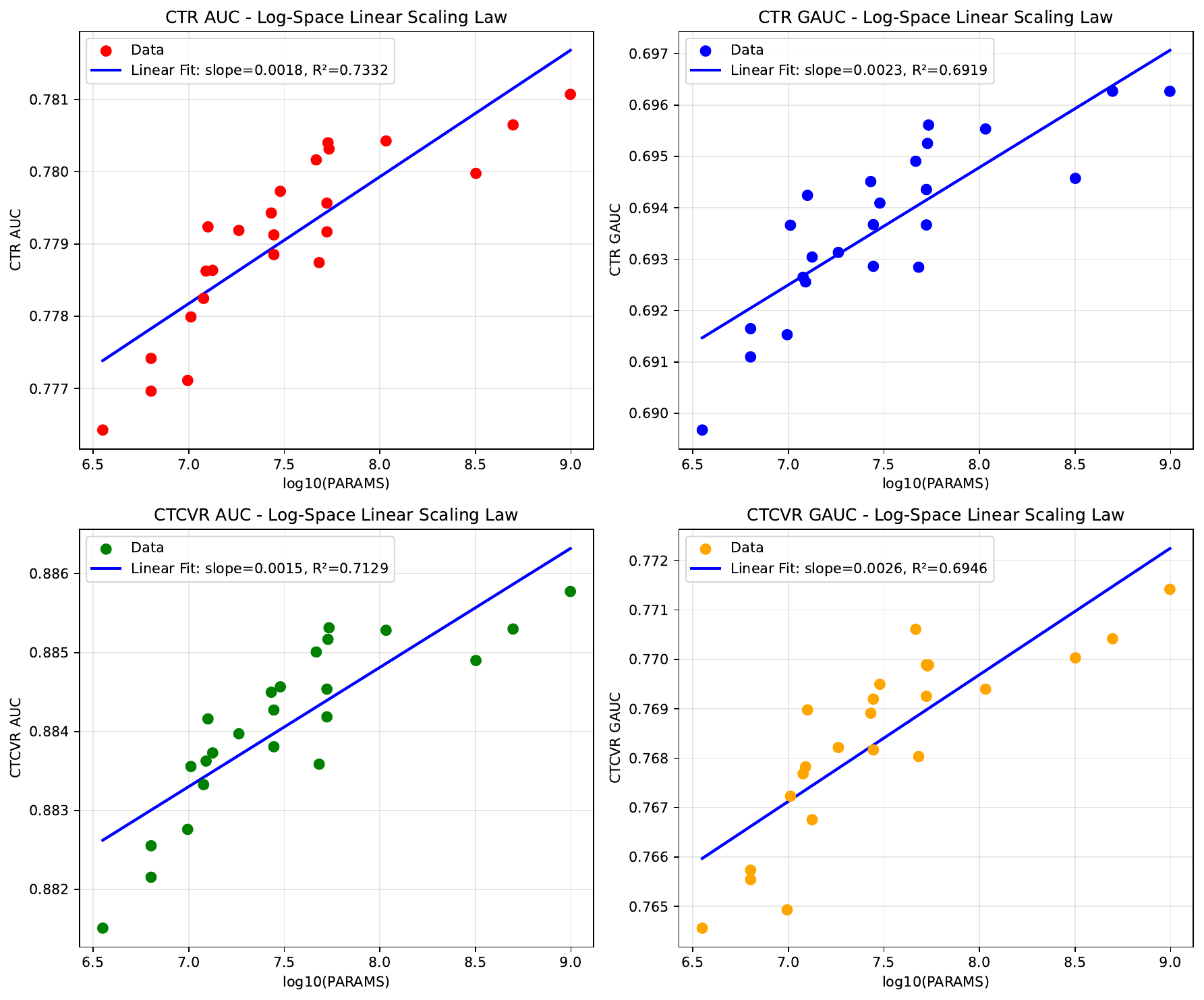}
\caption{Scaling laws between four metrics and
parameters of different models. All four metrics exhibit near-linear improvements with the logarithm of model size, confirming the power-law scaling behavior of MTmixAtt.}
\label{fig8}
\end{figure}

\textbf{Scaling Up.} Figure \ref{fig8} illustrates the power-law scaling behavior of our MTmixAtt model across four key metrics: CTR AUC, CTR GAUC, CTCVR AUC, and CTCVR GAUC. As the number of parameters increases from 15M to 1B, the model consistently improves on all evaluation metrics. This observation aligns with the log-linear scaling laws widely reported in large language models, 
suggesting that MTmixAtt preserves efficient capacity utilization as model size grows.

\subsection{Online A/B Test}
We further deployed MTmixAtt in Meituan’s online environment and conducted large-scale A/B testing to evaluate its effectiveness in real-world recommendation scenarios. As shown in Table \ref{tab:feed_metrics}, we not only verified its online performance in the \textit{Homepage} scenario, but also its online lift in the other three cross-scenarios. Specifically, Cross Scenarios 1–4 in the table correspond to \textit{Special Offer Groupon Feed}, \textit{Special Offer Groupon Top Card}, \textit{Deal Group Feed}, and \textit{Short Video} respectively.
In the \textit{Homepage} scenario, MTmixAtt increases Payment Page Views (PV) by +3.62\%, Actual Payment Gross Transaction Volume (GTV) by +2.54\%, and Novel Item Exposure Volume (NIEV) by +1.05\%. These consistent gains demonstrate that MTmixAtt enhances user engagement and directly contributes to transaction growth.
In other scenarios, MTmixAtt still delivers comprehensive improvements. For instance, in terms of Actual Payment GTV, it achieves lifts of 1.02\%, 2.71\%, 1.54\%, and 1.29\% respectively compared with the online baseline across the three cross-scenarios.

Overall, the online A/B tests confirm the practical value of MTmixAtt: it brings significant business improvements in both mainstream and specialized scenarios, validating its ability to scale effectively in large-scale industrial recommendation systems. These results highlight that MTmixAtt not only boosts commercial outcomes but also improves user experience by providing more relevant and diverse recommendations.

\section{Conclusion}
In this paper, we address the fundamental challenge of manual feature engineering and limited cross-scenario transfer in industrial recommendation systems. We propose MTmixAtt, a Mixture-of-Experts model with Multi-Mix Attention that unifies feature grouping, heterogeneous feature modeling, and cross-scenario adaptation within a single framework.
Our key contributions include: (1) an auto-group mechanism that automatically clusters arbitrary features into coherent tokens, eliminating manual feature engineering; (2) a novel expert architecture combining shared dense experts with scenario-aware sparse experts for effective cross-scenario modeling; and (3) a scalable framework that captures cross-token common patterns while preserving scenario-specific behaviors.
Extensive offline experiments on the industrial TRec dataset and large-scale online A/B tests on Meituan demonstrate that MTmixAtt consistently outperforms state-of-the-art baselines and delivers significant business gains. In future work, we plan to explore more efficient scaling strategies, extend MTmixAtt to multimodal recommendation settings, and investigate its applicability to other industrial domains beyond recommendation.

\appendix

\balance
\bibliographystyle{ACM-Reference-Format}

\bibliography{sample-base}


\begin{thebibliography}{34}


\ifx \showCODEN    \undefined \def \showCODEN     #1{\unskip}     \fi
\ifx \showISBNx    \undefined \def \showISBNx     #1{\unskip}     \fi
\ifx \showISBNxiii \undefined \def \showISBNxiii  #1{\unskip}     \fi
\ifx \showISSN     \undefined \def \showISSN      #1{\unskip}     \fi
\ifx \showLCCN     \undefined \def \showLCCN      #1{\unskip}     \fi
\ifx \shownote     \undefined \def \shownote      #1{#1}          \fi
\ifx \showarticletitle \undefined \def \showarticletitle #1{#1}   \fi
\ifx \showURL      \undefined \def \showURL       {\relax}        \fi
\providecommand\bibfield[2]{#2}
\providecommand\bibinfo[2]{#2}
\providecommand\natexlab[1]{#1}
\providecommand\showeprint[2][]{arXiv:#2}

\bibitem[Borisyuk et~al\mbox{.}(2024)]%
        {borisyuk2024lirankindustriallargescale}
\bibfield{author}{\bibinfo{person}{Fedor Borisyuk}, \bibinfo{person}{Mingzhou Zhou}, \bibinfo{person}{Qingquan Song}, \bibinfo{person}{Siyu Zhu}, \bibinfo{person}{Birjodh Tiwana}, \bibinfo{person}{Ganesh Parameswaran}, \bibinfo{person}{Siddharth Dangi}, \bibinfo{person}{Lars Hertel}, \bibinfo{person}{Qiang Xiao}, \bibinfo{person}{Xiaochen Hou}, \bibinfo{person}{Yunbo Ouyang}, \bibinfo{person}{Aman Gupta}, \bibinfo{person}{Sheallika Singh}, \bibinfo{person}{Dan Liu}, \bibinfo{person}{Hailing Cheng}, \bibinfo{person}{Lei Le}, \bibinfo{person}{Jonathan Hung}, \bibinfo{person}{Sathiya Keerthi}, \bibinfo{person}{Ruoyan Wang}, \bibinfo{person}{Fengyu Zhang}, \bibinfo{person}{Mohit Kothari}, \bibinfo{person}{Chen Zhu}, \bibinfo{person}{Daqi Sun}, \bibinfo{person}{Yun Dai}, \bibinfo{person}{Xun Luan}, \bibinfo{person}{Sirou Zhu}, \bibinfo{person}{Zhiwei Wang}, \bibinfo{person}{Neil Daftary}, \bibinfo{person}{Qianqi Shen}, \bibinfo{person}{Chengming Jiang}, \bibinfo{person}{Haichao Wei}, \bibinfo{person}{Maneesh
  Varshney}, \bibinfo{person}{Amol Ghoting}, {and} \bibinfo{person}{Souvik Ghosh}.} \bibinfo{year}{2024}\natexlab{}.
\newblock \bibinfo{title}{LiRank: Industrial Large Scale Ranking Models at LinkedIn}.
\newblock
\showeprint[arxiv]{2402.06859}~[cs.LG]
\urldef\tempurl%
\url{https://arxiv.org/abs/2402.06859}
\showURL{%
\tempurl}


\bibitem[Chai et~al\mbox{.}(2025)]%
        {chai2025longer}
\bibfield{author}{\bibinfo{person}{Zheng Chai}, \bibinfo{person}{Qin Ren}, \bibinfo{person}{Xijun Xiao}, \bibinfo{person}{Huizhi Yang}, \bibinfo{person}{Bo Han}, \bibinfo{person}{Sijun Zhang}, \bibinfo{person}{Di Chen}, \bibinfo{person}{Hui Lu}, \bibinfo{person}{Wenlin Zhao}, \bibinfo{person}{Lele Yu}, {et~al\mbox{.}}} \bibinfo{year}{2025}\natexlab{}.
\newblock \showarticletitle{Longer: Scaling up long sequence modeling in industrial recommenders}. In \bibinfo{booktitle}{\emph{Proceedings of the Nineteenth ACM Conference on Recommender Systems}}. \bibinfo{pages}{247--256}.
\newblock


\bibitem[Chang et~al\mbox{.}(2025)]%
        {ff5}
\bibfield{author}{\bibinfo{person}{Daryl Chang}, \bibinfo{person}{Yi Wu}, \bibinfo{person}{Jennifer She}, \bibinfo{person}{Li Wei}, {and} \bibinfo{person}{Lukasz Heldt}.} \bibinfo{year}{2025}\natexlab{}.
\newblock \bibinfo{title}{ACT: Automated Constraint Targeting for Multi-Objective Recommender Systems}.
\newblock
\showeprint[arxiv]{2509.03661}~[cs.IR]
\urldef\tempurl%
\url{https://arxiv.org/abs/2509.03661}
\showURL{%
\tempurl}


\bibitem[Chang et~al\mbox{.}(2023)]%
        {chang2023pepnet}
\bibfield{author}{\bibinfo{person}{Jianxin Chang}, \bibinfo{person}{Chenbin Zhang}, \bibinfo{person}{Yiqun Hui}, \bibinfo{person}{Dewei Leng}, \bibinfo{person}{Yanan Niu}, \bibinfo{person}{Yang Song}, {and} \bibinfo{person}{Kun Gai}.} \bibinfo{year}{2023}\natexlab{}.
\newblock \showarticletitle{Pepnet: Parameter and embedding personalized network for infusing with personalized prior information}. In \bibinfo{booktitle}{\emph{Proceedings of the 29th ACM SIGKDD Conference on Knowledge Discovery and Data Mining}}. \bibinfo{pages}{3795--3804}.
\newblock


\bibitem[Cheng et~al\mbox{.}(2016)]%
        {cheng2016widedeeplearning}
\bibfield{author}{\bibinfo{person}{Heng-Tze Cheng}, \bibinfo{person}{Levent Koc}, \bibinfo{person}{Jeremiah Harmsen}, \bibinfo{person}{Tal Shaked}, \bibinfo{person}{Tushar Chandra}, \bibinfo{person}{Hrishi Aradhye}, \bibinfo{person}{Glen Anderson}, \bibinfo{person}{Greg Corrado}, \bibinfo{person}{Wei Chai}, \bibinfo{person}{Mustafa Ispir}, \bibinfo{person}{Rohan Anil}, \bibinfo{person}{Zakaria Haque}, \bibinfo{person}{Lichan Hong}, \bibinfo{person}{Vihan Jain}, \bibinfo{person}{Xiaobing Liu}, {and} \bibinfo{person}{Hemal Shah}.} \bibinfo{year}{2016}\natexlab{}.
\newblock \bibinfo{title}{Wide \& Deep Learning for Recommender Systems}.
\newblock
\showeprint[arxiv]{1606.07792}~[cs.LG]
\urldef\tempurl%
\url{https://arxiv.org/abs/1606.07792}
\showURL{%
\tempurl}


\bibitem[Cokbas et~al\mbox{.}(2025)]%
        {ff1}
\bibfield{author}{\bibinfo{person}{Mertcan Cokbas}, \bibinfo{person}{Ziteng Liu}, \bibinfo{person}{Zeyi Tao}, \bibinfo{person}{Chengkai Zhang}, \bibinfo{person}{Elder Veliz}, \bibinfo{person}{Qin Huang}, \bibinfo{person}{Ellie Wen}, \bibinfo{person}{Huayu Li}, \bibinfo{person}{Qiang Jin}, \bibinfo{person}{Murat Duman}, \bibinfo{person}{Benjamin Au}, \bibinfo{person}{Guy Lebanon}, {and} \bibinfo{person}{Sagar Chordia}.} \bibinfo{year}{2025}\natexlab{}.
\newblock \bibinfo{title}{C2AL: Cohort-Contrastive Auxiliary Learning for Large-scale Recommendation Systems}.
\newblock
\showeprint[arxiv]{2510.02215}~[cs.LG]
\urldef\tempurl%
\url{https://arxiv.org/abs/2510.02215}
\showURL{%
\tempurl}


\bibitem[Dai et~al\mbox{.}(2024)]%
        {dai2024deepseekmoeultimateexpertspecialization}
\bibfield{author}{\bibinfo{person}{Damai Dai}, \bibinfo{person}{Chengqi Deng}, \bibinfo{person}{Chenggang Zhao}, \bibinfo{person}{R.~X. Xu}, \bibinfo{person}{Huazuo Gao}, \bibinfo{person}{Deli Chen}, \bibinfo{person}{Jiashi Li}, \bibinfo{person}{Wangding Zeng}, \bibinfo{person}{Xingkai Yu}, \bibinfo{person}{Y. Wu}, \bibinfo{person}{Zhenda Xie}, \bibinfo{person}{Y.~K. Li}, \bibinfo{person}{Panpan Huang}, \bibinfo{person}{Fuli Luo}, \bibinfo{person}{Chong Ruan}, \bibinfo{person}{Zhifang Sui}, {and} \bibinfo{person}{Wenfeng Liang}.} \bibinfo{year}{2024}\natexlab{}.
\newblock \bibinfo{title}{DeepSeekMoE: Towards Ultimate Expert Specialization in Mixture-of-Experts Language Models}.
\newblock
\showeprint[arxiv]{2401.06066}~[cs.CL]
\urldef\tempurl%
\url{https://arxiv.org/abs/2401.06066}
\showURL{%
\tempurl}


\bibitem[Dai et~al\mbox{.}(2025)]%
        {ff3}
\bibfield{author}{\bibinfo{person}{Sunhao Dai}, \bibinfo{person}{Jiakai Tang}, \bibinfo{person}{Jiahua Wu}, \bibinfo{person}{Kun Wang}, \bibinfo{person}{Yuxuan Zhu}, \bibinfo{person}{Bingjun Chen}, \bibinfo{person}{Bangyang Hong}, \bibinfo{person}{Yu Zhao}, \bibinfo{person}{Cong Fu}, \bibinfo{person}{Kangle Wu}, \bibinfo{person}{Yabo Ni}, \bibinfo{person}{Anxiang Zeng}, \bibinfo{person}{Wenjie Wang}, \bibinfo{person}{Xu Chen}, \bibinfo{person}{Jun Xu}, {and} \bibinfo{person}{See-Kiong Ng}.} \bibinfo{year}{2025}\natexlab{}.
\newblock \bibinfo{title}{OnePiece: Bringing Context Engineering and Reasoning to Industrial Cascade Ranking System}.
\newblock
\showeprint[arxiv]{2509.18091}~[cs.IR]
\urldef\tempurl%
\url{https://arxiv.org/abs/2509.18091}
\showURL{%
\tempurl}


\bibitem[DeepSeek-AI et~al\mbox{.}(2025)]%
        {deepseekai2025deepseekv3technicalreport}
\bibfield{author}{\bibinfo{person}{DeepSeek-AI}, \bibinfo{person}{Aixin Liu}, \bibinfo{person}{Bei Feng}, \bibinfo{person}{Bing Xue}, \bibinfo{person}{Bingxuan Wang}, \bibinfo{person}{Bochao Wu}, {and} \bibinfo{person}{Chengda Lu}.} \bibinfo{year}{2025}\natexlab{}.
\newblock \bibinfo{title}{DeepSeek-V3 Technical Report}.
\newblock
\showeprint[arxiv]{2412.19437}~[cs.CL]
\urldef\tempurl%
\url{https://arxiv.org/abs/2412.19437}
\showURL{%
\tempurl}


\bibitem[Gheewala et~al\mbox{.}(2025)]%
        {Ref2}
\bibfield{author}{\bibinfo{person}{Shivangi Gheewala}, \bibinfo{person}{Shuxiang Xu}, {and} \bibinfo{person}{Soonja Yeom}.} \bibinfo{year}{2025}\natexlab{}.
\newblock \showarticletitle{In-depth survey: deep learning in recommender systems—exploring prediction and ranking models, datasets, feature analysis, and emerging trends}.
\newblock \bibinfo{journal}{\emph{Neural Computing and Applications}}  \bibinfo{volume}{37} (\bibinfo{date}{03} \bibinfo{year}{2025}), \bibinfo{pages}{10875--10947}.
\newblock
\href{https://doi.org/10.1007/s00521-024-10866-z}{doi:\nolinkurl{10.1007/s00521-024-10866-z}}


\bibitem[Gui et~al\mbox{.}(2023)]%
        {hiformer}
\bibfield{author}{\bibinfo{person}{Huan Gui}, \bibinfo{person}{Ruoxi Wang}, \bibinfo{person}{Ke Yin}, \bibinfo{person}{Long Jin}, \bibinfo{person}{Maciej Kula}, \bibinfo{person}{Taibai Xu}, \bibinfo{person}{Lichan Hong}, {and} \bibinfo{person}{Ed~H. Chi}.} \bibinfo{year}{2023}\natexlab{}.
\newblock \bibinfo{title}{Hiformer: Heterogeneous Feature Interactions Learning with Transformers for Recommender Systems}.
\newblock
\showeprint[arxiv]{2311.05884}~[cs.IR]
\urldef\tempurl%
\url{https://arxiv.org/abs/2311.05884}
\showURL{%
\tempurl}


\bibitem[Guo et~al\mbox{.}(2018)]%
        {guo2018deepfmendtoendwide}
\bibfield{author}{\bibinfo{person}{Huifeng Guo}, \bibinfo{person}{Ruiming Tang}, \bibinfo{person}{Yunming Ye}, \bibinfo{person}{Zhenguo Li}, \bibinfo{person}{Xiuqiang He}, {and} \bibinfo{person}{Zhenhua Dong}.} \bibinfo{year}{2018}\natexlab{}.
\newblock \bibinfo{title}{DeepFM: An End-to-End Wide \& Deep Learning Framework for CTR Prediction}.
\newblock
\showeprint[arxiv]{1804.04950}~[cs.IR]
\urldef\tempurl%
\url{https://arxiv.org/abs/1804.04950}
\showURL{%
\tempurl}


\bibitem[Hu et~al\mbox{.}(2025)]%
        {hu2025dynamicforgettingspatiotemporalperiodic}
\bibfield{author}{\bibinfo{person}{Zhaoyu Hu}, \bibinfo{person}{Hao Guo}, \bibinfo{person}{Yuan Tian}, \bibinfo{person}{Erpeng Xue}, \bibinfo{person}{Jianyang Wang}, \bibinfo{person}{Xianyang Qi}, \bibinfo{person}{Hongxiang Lin}, \bibinfo{person}{Lei Wang}, {and} \bibinfo{person}{Sheng Chen}.} \bibinfo{year}{2025}\natexlab{}.
\newblock \bibinfo{title}{Dynamic Forgetting and Spatio-Temporal Periodic Interest Modeling for Local-Life Service Recommendation}.
\newblock
\showeprint[arxiv]{2508.02451}~[cs.IR]
\urldef\tempurl%
\url{https://arxiv.org/abs/2508.02451}
\showURL{%
\tempurl}


\bibitem[Lan et~al\mbox{.}(2025)]%
        {lan2025nextuserretrievalenhancingcoldstart}
\bibfield{author}{\bibinfo{person}{Yu-Ting Lan}, \bibinfo{person}{Yang Huo}, \bibinfo{person}{Yi Shen}, \bibinfo{person}{Xiao Yang}, {and} \bibinfo{person}{Zuotao Liu}.} \bibinfo{year}{2025}\natexlab{}.
\newblock \bibinfo{title}{Next-User Retrieval: Enhancing Cold-Start Recommendations via Generative Next-User Modeling}.
\newblock
\showeprint[arxiv]{2506.15267}~[cs.IR]
\urldef\tempurl%
\url{https://arxiv.org/abs/2506.15267}
\showURL{%
\tempurl}


\bibitem[Li et~al\mbox{.}(2024)]%
        {multicene}
\bibfield{author}{\bibinfo{person}{Wenhao Li}, \bibinfo{person}{Jie Zhou}, \bibinfo{person}{Chuan Luo}, \bibinfo{person}{Chao Tang}, \bibinfo{person}{Kun Zhang}, {and} \bibinfo{person}{Shixiong Zhao}.} \bibinfo{year}{2024}\natexlab{}.
\newblock \showarticletitle{Scene-wise Adaptive Network for Dynamic Cold-start Scenes Optimization in CTR Prediction}. In \bibinfo{booktitle}{\emph{Proceedings of the 18th ACM Conference on Recommender Systems}} (Bari, Italy) \emph{(\bibinfo{series}{RecSys '24})}. \bibinfo{publisher}{Association for Computing Machinery}, \bibinfo{address}{New York, NY, USA}, \bibinfo{pages}{370–379}.
\newblock
\showISBNx{9798400705052}
\href{https://doi.org/10.1145/3640457.3688115}{doi:\nolinkurl{10.1145/3640457.3688115}}


\bibitem[Lin et~al\mbox{.}(2025)]%
        {lin2025onlinegradientboostingdecision}
\bibfield{author}{\bibinfo{person}{Huawei Lin}, \bibinfo{person}{Jun~Woo Chung}, \bibinfo{person}{Yingjie Lao}, {and} \bibinfo{person}{Weijie Zhao}.} \bibinfo{year}{2025}\natexlab{}.
\newblock \bibinfo{title}{Online Gradient Boosting Decision Tree: In-Place Updates for Efficient Adding/Deleting Data}.
\newblock
\showeprint[arxiv]{2502.01634}~[cs.LG]
\urldef\tempurl%
\url{https://arxiv.org/abs/2502.01634}
\showURL{%
\tempurl}


\bibitem[Liu et~al\mbox{.}(2025)]%
        {ff2}
\bibfield{author}{\bibinfo{person}{Jingzhe Liu}, \bibinfo{person}{Liam Collins}, \bibinfo{person}{Jiliang Tang}, \bibinfo{person}{Tong Zhao}, \bibinfo{person}{Neil Shah}, {and} \bibinfo{person}{Clark~Mingxuan Ju}.} \bibinfo{year}{2025}\natexlab{}.
\newblock \bibinfo{title}{Understanding Generative Recommendation with Semantic IDs from a Model-scaling View}.
\newblock
\showeprint[arxiv]{2509.25522}~[cs.AI]
\urldef\tempurl%
\url{https://arxiv.org/abs/2509.25522}
\showURL{%
\tempurl}


\bibitem[Liu et~al\mbox{.}(2019)]%
        {bu1}
\bibfield{author}{\bibinfo{person}{Tianqiao Liu}, \bibinfo{person}{Zhiwei Wang}, \bibinfo{person}{Jiliang Tang}, \bibinfo{person}{Songfan Yang}, \bibinfo{person}{Gale~Yan Huang}, {and} \bibinfo{person}{Zitao Liu}.} \bibinfo{year}{2019}\natexlab{}.
\newblock \showarticletitle{Recommender Systems with Heterogeneous Side Information}. In \bibinfo{booktitle}{\emph{The World Wide Web Conference}} (San Francisco, CA, USA) \emph{(\bibinfo{series}{WWW '19})}. \bibinfo{publisher}{Association for Computing Machinery}, \bibinfo{address}{New York, NY, USA}, \bibinfo{pages}{3027–3033}.
\newblock
\showISBNx{9781450366748}
\href{https://doi.org/10.1145/3308558.3313580}{doi:\nolinkurl{10.1145/3308558.3313580}}


\bibitem[Liu et~al\mbox{.}(2024)]%
        {GNNs}
\bibfield{author}{\bibinfo{person}{Yuxi Liu}, \bibinfo{person}{Lianghao Xia}, {and} \bibinfo{person}{Chao Huang}.} \bibinfo{year}{2024}\natexlab{}.
\newblock \showarticletitle{SelfGNN: Self-Supervised Graph Neural Networks for Sequential Recommendation}. In \bibinfo{booktitle}{\emph{Proceedings of the 47th International ACM SIGIR Conference on Research and Development in Information Retrieval}} (Washington DC, USA) \emph{(\bibinfo{series}{SIGIR '24})}. \bibinfo{publisher}{Association for Computing Machinery}, \bibinfo{address}{New York, NY, USA}, \bibinfo{pages}{1609–1618}.
\newblock
\showISBNx{9798400704314}
\href{https://doi.org/10.1145/3626772.3657716}{doi:\nolinkurl{10.1145/3626772.3657716}}


\bibitem[Qu et~al\mbox{.}(2016)]%
        {PNN}
\bibfield{author}{\bibinfo{person}{Yanru Qu}, \bibinfo{person}{Han Cai}, \bibinfo{person}{Kan Ren}, \bibinfo{person}{Weinan Zhang}, \bibinfo{person}{Yong Yu}, \bibinfo{person}{Ying Wen}, {and} \bibinfo{person}{Jun Wang}.} \bibinfo{year}{2016}\natexlab{}.
\newblock \bibinfo{title}{Product-based Neural Networks for User Response Prediction}.
\newblock
\showeprint[arxiv]{1611.00144}~[cs.LG]
\urldef\tempurl%
\url{https://arxiv.org/abs/1611.00144}
\showURL{%
\tempurl}


\bibitem[Rendle(2010)]%
        {FM}
\bibfield{author}{\bibinfo{person}{Steffen Rendle}.} \bibinfo{year}{2010}\natexlab{}.
\newblock \showarticletitle{Factorization Machines}. In \bibinfo{booktitle}{\emph{2010 IEEE International Conference on Data Mining}}. \bibinfo{pages}{995--1000}.
\newblock
\href{https://doi.org/10.1109/ICDM.2010.127}{doi:\nolinkurl{10.1109/ICDM.2010.127}}


\bibitem[Song et~al\mbox{.}(2019)]%
        {Autoint}
\bibfield{author}{\bibinfo{person}{Weiping Song}, \bibinfo{person}{Chence Shi}, \bibinfo{person}{Zhiping Xiao}, \bibinfo{person}{Zhijian Duan}, \bibinfo{person}{Yewen Xu}, \bibinfo{person}{Ming Zhang}, {and} \bibinfo{person}{Jian Tang}.} \bibinfo{year}{2019}\natexlab{}.
\newblock \showarticletitle{AutoInt: Automatic Feature Interaction Learning via Self-Attentive Neural Networks}. In \bibinfo{booktitle}{\emph{Proceedings of the 28th ACM International Conference on Information and Knowledge Management}} \emph{(\bibinfo{series}{CIKM ’19})}. \bibinfo{publisher}{ACM}, \bibinfo{pages}{1161–1170}.
\newblock
\href{https://doi.org/10.1145/3357384.3357925}{doi:\nolinkurl{10.1145/3357384.3357925}}


\bibitem[Su et~al\mbox{.}(2022)]%
        {Su_2022}
\bibfield{author}{\bibinfo{person}{Yixin Su}, \bibinfo{person}{Yunxiang Zhao}, \bibinfo{person}{Sarah Erfani}, \bibinfo{person}{Junhao Gan}, {and} \bibinfo{person}{Rui Zhang}.} \bibinfo{year}{2022}\natexlab{}.
\newblock \showarticletitle{Detecting Arbitrary Order Beneficial Feature Interactions for Recommender Systems}. In \bibinfo{booktitle}{\emph{Proceedings of the 28th ACM SIGKDD Conference on Knowledge Discovery and Data Mining}} \emph{(\bibinfo{series}{KDD ’22})}. \bibinfo{publisher}{ACM}, \bibinfo{pages}{1676–1686}.
\newblock
\href{https://doi.org/10.1145/3534678.3539238}{doi:\nolinkurl{10.1145/3534678.3539238}}


\bibitem[Team et~al\mbox{.}(2025)]%
        {kimiteam2025kimik2openagentic}
\bibfield{author}{\bibinfo{person}{Kimi Team}, \bibinfo{person}{Yifan Bai}, \bibinfo{person}{Yiping Bao}, \bibinfo{person}{Guanduo Chen}, {and} \bibinfo{person}{Jiahao Chen}.} \bibinfo{year}{2025}\natexlab{}.
\newblock \bibinfo{title}{Kimi K2: Open Agentic Intelligence}.
\newblock
\showeprint[arxiv]{2507.20534}~[cs.LG]
\urldef\tempurl%
\url{https://arxiv.org/abs/2507.20534}
\showURL{%
\tempurl}


\bibitem[Tian et~al\mbox{.}(2023)]%
        {tian2023multi}
\bibfield{author}{\bibinfo{person}{Yu Tian}, \bibinfo{person}{Bofang Li}, \bibinfo{person}{Si Chen}, \bibinfo{person}{Xubin Li}, \bibinfo{person}{Hongbo Deng}, \bibinfo{person}{Jian Xu}, \bibinfo{person}{Bo Zheng}, \bibinfo{person}{Qian Wang}, {and} \bibinfo{person}{Chenliang Li}.} \bibinfo{year}{2023}\natexlab{}.
\newblock \showarticletitle{Multi-scenario ranking with adaptive feature learning}. In \bibinfo{booktitle}{\emph{Proceedings of the 46th International ACM SIGIR Conference on Research and Development in Information Retrieval}}. \bibinfo{pages}{517--526}.
\newblock


\bibitem[Tolstikhin et~al\mbox{.}(2021)]%
        {MLP-Mixer}
\bibfield{author}{\bibinfo{person}{Ilya Tolstikhin}, \bibinfo{person}{Neil Houlsby}, \bibinfo{person}{Alexander Kolesnikov}, \bibinfo{person}{Lucas Beyer}, \bibinfo{person}{Xiaohua Zhai}, \bibinfo{person}{Thomas Unterthiner}, \bibinfo{person}{Jessica Yung}, \bibinfo{person}{Andreas Steiner}, \bibinfo{person}{Daniel Keysers}, \bibinfo{person}{Jakob Uszkoreit}, \bibinfo{person}{Mario Lucic}, {and} \bibinfo{person}{Alexey Dosovitskiy}.} \bibinfo{year}{2021}\natexlab{}.
\newblock \showarticletitle{MLP-mixer: an all-MLP architecture for vision}. In \bibinfo{booktitle}{\emph{Proceedings of the 35th International Conference on Neural Information Processing Systems}} \emph{(\bibinfo{series}{NIPS '21})}. \bibinfo{publisher}{Curran Associates Inc.}, \bibinfo{address}{Red Hook, NY, USA}, Article \bibinfo{articleno}{1857}, \bibinfo{numpages}{12}~pages.
\newblock
\showISBNx{9781713845393}


\bibitem[Wang et~al\mbox{.}(2021)]%
        {DCNv2}
\bibfield{author}{\bibinfo{person}{Ruoxi Wang}, \bibinfo{person}{Rakesh Shivanna}, \bibinfo{person}{Derek Cheng}, \bibinfo{person}{Sagar Jain}, \bibinfo{person}{Dong Lin}, \bibinfo{person}{Lichan Hong}, {and} \bibinfo{person}{Ed Chi}.} \bibinfo{year}{2021}\natexlab{}.
\newblock \showarticletitle{DCN V2: Improved Deep \& Cross Network and Practical Lessons for Web-scale Learning to Rank Systems}. In \bibinfo{booktitle}{\emph{Proceedings of the Web Conference 2021}} (Ljubljana, Slovenia) \emph{(\bibinfo{series}{WWW '21})}. \bibinfo{publisher}{Association for Computing Machinery}, \bibinfo{address}{New York, NY, USA}, \bibinfo{pages}{1785–1797}.
\newblock
\showISBNx{9781450383127}
\href{https://doi.org/10.1145/3442381.3450078}{doi:\nolinkurl{10.1145/3442381.3450078}}


\bibitem[Wang et~al\mbox{.}(2024)]%
        {wang2024homehierarchymultigateexperts}
\bibfield{author}{\bibinfo{person}{Xu Wang}, \bibinfo{person}{Jiangxia Cao}, \bibinfo{person}{Zhiyi Fu}, \bibinfo{person}{Kun Gai}, {and} \bibinfo{person}{Guorui Zhou}.} \bibinfo{year}{2024}\natexlab{}.
\newblock \bibinfo{title}{HoME: Hierarchy of Multi-Gate Experts for Multi-Task Learning at Kuaishou}.
\newblock
\showeprint[arxiv]{2408.05430}~[cs.IR]
\urldef\tempurl%
\url{https://arxiv.org/abs/2408.05430}
\showURL{%
\tempurl}


\bibitem[Wang et~al\mbox{.}(2025)]%
        {wang2025home}
\bibfield{author}{\bibinfo{person}{Xu Wang}, \bibinfo{person}{Jiangxia Cao}, \bibinfo{person}{Zhiyi Fu}, \bibinfo{person}{Kun Gai}, {and} \bibinfo{person}{Guorui Zhou}.} \bibinfo{year}{2025}\natexlab{}.
\newblock \showarticletitle{Home: Hierarchy of multi-gate experts for multi-task learning at kuaishou}. In \bibinfo{booktitle}{\emph{Proceedings of the 31st ACM SIGKDD Conference on Knowledge Discovery and Data Mining V. 1}}. \bibinfo{pages}{2638--2647}.
\newblock


\bibitem[Yang et~al\mbox{.}(2024)]%
        {yang2024mloramultidomainlowrankadaptive}
\bibfield{author}{\bibinfo{person}{Zhiming Yang}, \bibinfo{person}{Haining Gao}, \bibinfo{person}{Dehong Gao}, \bibinfo{person}{Luwei Yang}, \bibinfo{person}{Libin Yang}, \bibinfo{person}{Xiaoyan Cai}, \bibinfo{person}{Wei Ning}, {and} \bibinfo{person}{Guannan Zhang}.} \bibinfo{year}{2024}\natexlab{}.
\newblock \bibinfo{title}{MLoRA: Multi-Domain Low-Rank Adaptive Network for CTR Prediction}.
\newblock
\showeprint[arxiv]{2408.08913}~[cs.IR]
\urldef\tempurl%
\url{https://arxiv.org/abs/2408.08913}
\showURL{%
\tempurl}


\bibitem[Zhang et~al\mbox{.}(2024)]%
        {wukong}
\bibfield{author}{\bibinfo{person}{Buyun Zhang}, \bibinfo{person}{Liang Luo}, \bibinfo{person}{Yuxin Chen}, \bibinfo{person}{Jade Nie}, \bibinfo{person}{Xi Liu}, \bibinfo{person}{Daifeng Guo}, \bibinfo{person}{Yanli Zhao}, \bibinfo{person}{Shen Li}, \bibinfo{person}{Yuchen Hao}, \bibinfo{person}{Yantao Yao}, \bibinfo{person}{Guna Lakshminarayanan}, \bibinfo{person}{Ellie~Dingqiao Wen}, \bibinfo{person}{Jongsoo Park}, \bibinfo{person}{Maxim Naumov}, {and} \bibinfo{person}{Wenlin Chen}.} \bibinfo{year}{2024}\natexlab{}.
\newblock \bibinfo{title}{Wukong: Towards a Scaling Law for Large-Scale Recommendation}.
\newblock
\showeprint[arxiv]{2403.02545}~[cs.LG]
\urldef\tempurl%
\url{https://arxiv.org/abs/2403.02545}
\showURL{%
\tempurl}


\bibitem[Zhou et~al\mbox{.}(2023)]%
        {zhou2023hinet}
\bibfield{author}{\bibinfo{person}{Jie Zhou}, \bibinfo{person}{Xianshuai Cao}, \bibinfo{person}{Wenhao Li}, \bibinfo{person}{Lin Bo}, \bibinfo{person}{Kun Zhang}, \bibinfo{person}{Chuan Luo}, {and} \bibinfo{person}{Qian Yu}.} \bibinfo{year}{2023}\natexlab{}.
\newblock \showarticletitle{Hinet: Novel multi-scenario \& multi-task learning with hierarchical information extraction}. In \bibinfo{booktitle}{\emph{2023 IEEE 39th International Conference on Data Engineering (ICDE)}}. IEEE, \bibinfo{pages}{2969--2975}.
\newblock


\bibitem[Zhu et~al\mbox{.}(2025)]%
        {zhu2025rankmixerscalingrankingmodels}
\bibfield{author}{\bibinfo{person}{Jie Zhu}, \bibinfo{person}{Zhifang Fan}, \bibinfo{person}{Xiaoxie Zhu}, \bibinfo{person}{Yuchen Jiang}, \bibinfo{person}{Hangyu Wang}, \bibinfo{person}{Xintian Han}, \bibinfo{person}{Haoran Ding}, \bibinfo{person}{Xinmin Wang}, \bibinfo{person}{Wenlin Zhao}, \bibinfo{person}{Zhen Gong}, \bibinfo{person}{Huizhi Yang}, \bibinfo{person}{Zheng Chai}, \bibinfo{person}{Zhe Chen}, \bibinfo{person}{Yuchao Zheng}, \bibinfo{person}{Qiwei Chen}, \bibinfo{person}{Feng Zhang}, \bibinfo{person}{Xun Zhou}, \bibinfo{person}{Peng Xu}, \bibinfo{person}{Xiao Yang}, \bibinfo{person}{Di Wu}, {and} \bibinfo{person}{Zuotao Liu}.} \bibinfo{year}{2025}\natexlab{}.
\newblock \bibinfo{title}{RankMixer: Scaling Up Ranking Models in Industrial Recommenders}.
\newblock
\showeprint[arxiv]{2507.15551}~[cs.IR]
\urldef\tempurl%
\url{https://arxiv.org/abs/2507.15551}
\showURL{%
\tempurl}


\bibitem[Zou and Sun(2025)]%
        {Ref1}
\bibfield{author}{\bibinfo{person}{Kuan Zou} {and} \bibinfo{person}{Aixin Sun}.} \bibinfo{year}{2025}\natexlab{}.
\newblock \bibinfo{title}{A Survey of Real-World Recommender Systems: Challenges, Constraints, and Industrial Perspectives}.
\newblock
\showeprint[arxiv]{2509.06002}~[cs.IR]
\urldef\tempurl%
\url{https://arxiv.org/abs/2509.06002}
\showURL{%
\tempurl}


\end{thebibliography}
\end{document}